\documentclass[showpacs, oneside, twocolumn, prl, amsmath, amssymb, nofootinbib, superscriptaddress]{revtex4-1}

\usepackage{cases}
\usepackage{amsmath}
\usepackage{amssymb}
\usepackage{amsfonts}
\usepackage{amssymb}
\usepackage{dcolumn}
\usepackage{bm}
\usepackage{bbm}
\usepackage{graphicx}
\usepackage{xcolor}
\usepackage{array}
\usepackage{subfigure}
\usepackage{hyperref}

\newcommand{\be}{\begin{equation}}
\newcommand{\ee}{\end{equation}}
\newcommand{\ba}{\begin{eqnarray}}
\newcommand{\ea}{\end{eqnarray}}
\newcommand{\no}{\nonumber \\}
\newcommand{\gsim}{\mathrel{\hbox{\rlap{\lower.55ex \hbox {$\sim$}}
                   \kern-.3em \raise.4ex \hbox{$>$}}}}
\newcommand{\lsim}{\mathrel{\hbox{\rlap{\lower.55ex \hbox {$\sim$}}
                   \kern-.3em \raise.4ex \hbox{$<$}}}}

\def\roughly#1{\mathrel{\raise.3ex\hbox{$#1$\kern-.75em%
\lower1ex\hbox{$\sim$}}}}
\def\lsim{\roughly<}
\def\gsim{\roughly>}

\def\({\left(}
\def\){\right)}
\def\[{\left[}
\def\]{\right]}
\def\<{\langle}
\def\>{\rangle}
\def\pd{\partial}

\def\L{{\Lambda}}
\def\d{{\delta}}
\def\D{{\Delta}}
\def\o{{\omega}}

\def\h{{\eta}}

\def\m{{\mu}}

\def\r{{\rho}}
\def\s{{\sigma}}

\def\t{{\tau}}

\def\Ph{{\Phi}}
\def\Ps{{\Psi}}

\newcommand{\cN}{{\cal N}}
\newcommand{\cA}{{\cal A}}

\hypersetup{colorlinks=true,
            breaklinks=true,
            pdfstartview=Fit,
            linkcolor=blue,
            citecolor=blue,
            urlcolor=blue}

\begin{document}

\title{Holographic Preheating}

\author{Yi-Fu Cai}
\email{yifucai@ustc.edu.cn}
\affiliation{CAS Key Laboratory for Research in Galaxies and Cosmology, Department of Astronomy, \\
University of Science and Technology of China, Chinese Academy of Sciences, Hefei, Anhui 230026, China}

\author{Shu Lin}
\email{linshu8@mail.sysu.edu.cn}
\affiliation{School of Physics and Astronomy, Sun Yat-Sen University, Guangzhou 510275, China}

\author{Junyu Liu}
\email{jliu2@caltech.edu}
\affiliation{CAS Key Laboratory for Research in Galaxies and Cosmology, Department of Astronomy, \\
University of Science and Technology of China, Chinese Academy of Sciences, Hefei, Anhui 230026, China}
\affiliation{Walter Burke Institute for Theoretical Physics, California Institute of Technology, Pasadena, CA 91125, United States}

\author{Jia-Rui Sun}
\email{sunjiarui@sysu.edu.cn}
\affiliation{School of Physics and Astronomy, Sun Yat-Sen University, Guangzhou 510275, China}

\begin{abstract}
We propose a holographic description of cosmic preheating at strong coupling.
In this scenario the energy transfer between the inflaton and matter field is mimicked by a model of holographic superconductor.
An exponential amplification of the matter field during preheating can be described by the quasi-normal modes of a metastable ``black hole'' in the bulk spacetime with an expanding boundary.
Our results reveal that the matter field can be produced continuously at strong coupling in contrast to the case of weak coupling with a discontinuous matter growth as inflaton oscillates.
Furthermore, the amplification of matter field has an enhanced dependence on the vacuum expectation value of the inflaton at strong coupling.
By virtue of the proposed mechanism, physics of the very early universe at an extremely high temperature right after inflation may become accessible.
\end{abstract}

\pacs{98.80.Cq, 11.25.Tq, 74.20.-z, 04.50.Gh}

\maketitle

%\section{Introduction}
{\it Introduction} --
Cosmic reheating, which bridges the gap between the end of the primordial inflationary universe
and the beginning of the big bang universe, is one significant puzzle in modern cosmology.
It is suggested that after inflation, the energy restored in the potential of the inflaton field can be released by producing matter fields in the standard model (SM) of particle physics through their couplings.
The reheating was studied using the description of the lowest order perturbation theory in \cite{Abbott:1982hn, Dolgov:1982th, Albrecht:1982mp} and cosmologists found that this process was typically slow and would lead to an initial state of low temperature for the hot big bang. As was observed in \cite{Dolgov:1989us, Traschen:1990sw}, however, an instability of parametric resonance may exist and realize a drastic phase of energy transfer from the inflaton to matter fields, which the inflaton weakly couples to. This phase, dubbed as {\it preheating}, was studied extensively in the literature \cite{Kofman:1994rk, Shtanov:1994ce, Kofman:1997yn} (see e.g. \cite{Bassett:2005xm, Allahverdi:2010xz, Amin:2014eta} for recent reviews; and see \cite{Cai:2011ci} for preheating in bounce cosmology). The instability of parametric amplification may also occur in metric fluctuations, and hence, the study of preheating can play a crucial role to observationally constrain or even rule out inflationary models \cite{Finelli:2000ya, Bassett:1999cg, Brandenberger:2007ca, Brandenberger:2008if, Moghaddam:2014ksa, McDonough:2016xvu}.

The perturbative analysis indicates that the parametric resonance, characterized by a period of exponential growth of the matter fields, terminates when the backreaction is non-negligible. Preheating is followed by the continued matter field evolution and its equilibration through the energy redistribution until the universe reaches thermal equilibrium \cite{Micha:2002ey, Micha:2004bv}. This completes the reheating process and marks the beginning of the hot big bang. Interestingly, an analogous scenario also arises in the context of heavy ion collisions, where the physical system evolves from saturated nucleus to quark gluon plasma in local thermal equilibrium \cite{Berges:2008pc}.
The above scenario is well established upon the assumption that the inflaton weakly couples to matter fields. Many features characteristic of the weakly coupled theory follows naturally from the assumption, e.g. the SM particles are produced only when the inflaton rolls over the bottom of potential well. While the nature of the inflaton field and its couplings to matter fields remain mystery, it is desirable to explore the preheating process in a strongly coupled theory as an alternative. In addressing this issue, we can also gain insights on which features are generic for cosmic preheating and which features are more specific to underlying models.

\begin{figure}
\begin{center}
\includegraphics[width=0.4\textwidth]{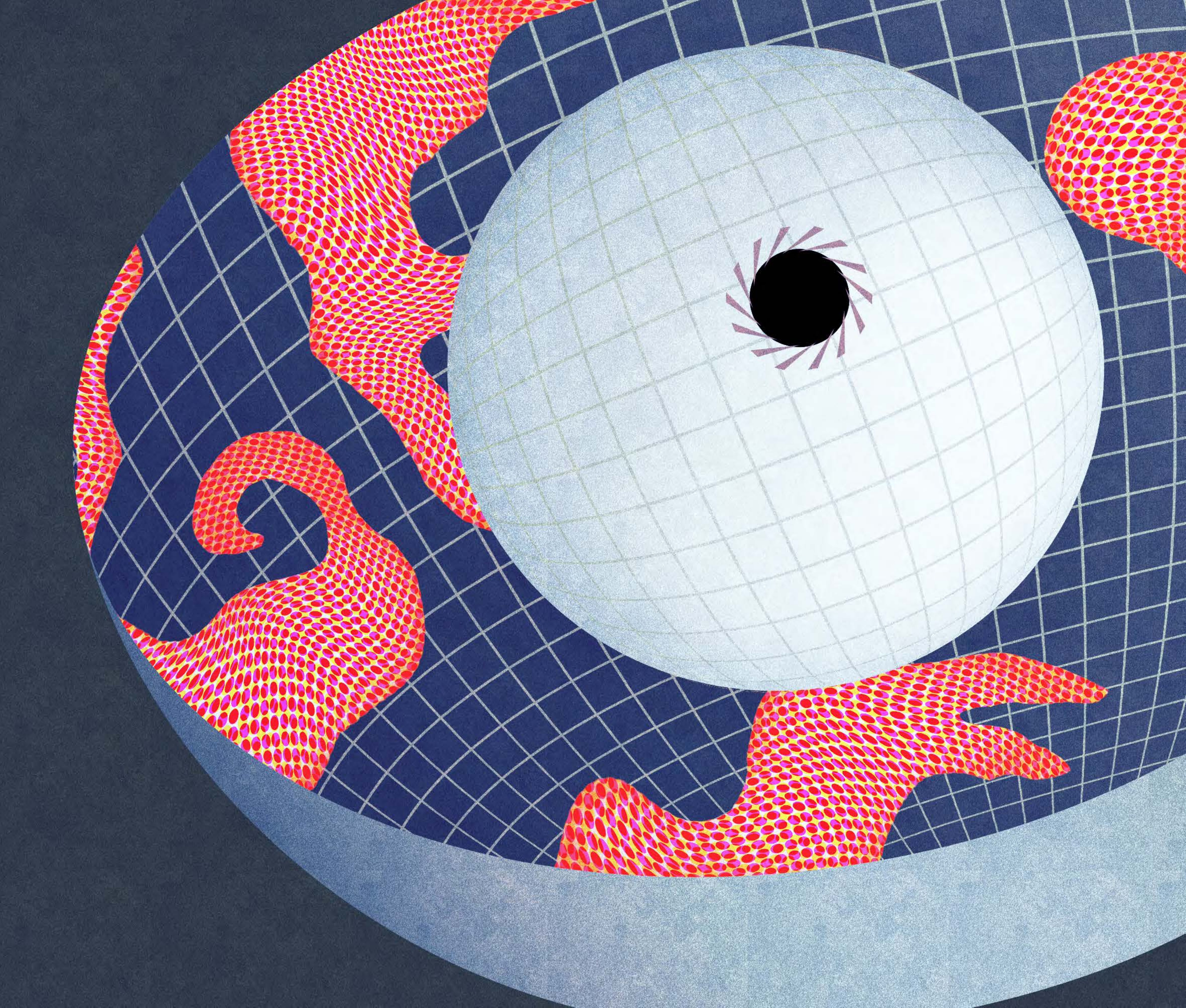}
\end{center}
\caption{Artistic illustration of {\it holographic preheating}. A 4D FRW universe is holographically described by a 5D AdS-Schwarzschild black hole, which is conformally mapped to an AdS-FRW background. (credit to Mr. Yulin Hu)
}\label{illustration}
\end{figure}

For pedagogical purposes, we artistically illustrate a holographic description of preheating at strong coupling in Fig.~\ref{illustration}.
We use a metastable hairy black hole as an initial state. Its evolution towards a stable black hole can mimic the preheating process.
Since preheating occurs in an expanding universe, we apply a holographic model with a Friedmann-Robertson-Walker (FRW) boundary
\footnote{Note that, our starting point differs from other holographic descriptions of cosmology where the universe itself is a bulk spacetime with a field theory boundary, such as, involving a time-varying gravitational coupling \cite{Chu:2006pa, Das:2006dz, Das:2006pw, Awad:2009bh}, in a nontrivial unstable state \cite{Hertog:2005hu, Craps:2007ch}, or, including field fluctuations \cite{Brandenberger:2016egn, Ferreira:2016gfg}. The reheating through holographic thermalization was studied in \cite{Kawai:2015lja}.}.
The bulk spacetime is then an Anti de Sitter (AdS)-FRW and such a background can be conformally mapped to the familiar AdS-Schwarzschild black hole as analyzed in \cite{Apostolopoulos:2008ru}.
In order to extract the detailed information, one can proceed with the calculation of quasi-normal modes (QNMs) in this background.
The QNMs contain an unstable mode, which exactly characterizes the exponential matter creation during preheating at strong coupling.
%And the QNMs are conformally related to the unstable modes in the FRW boundary, which correspond to the exponential matter creations in the mimicked universe.

%We suggest that a model of holographic superconductor can be used to analyze the detailed information about cosmic preheating at strong coupling.
%Since preheating occurs in an expanding universe, we apply a holographic model with an Friedmann-Robertson-Walker (FRW) boundary\footnote{Note that, our starting point differs from other holographic descriptions of the very early universe in which the universe itself is a bulk spacetime with a field theory boundary, such as, involving a time-varying gravitational coupling \cite{Chu:2006pa, Das:2006dz, Das:2006pw, Awad:2009bh}, in a nontrivial unstable state \cite{Hertog:2005hu, Craps:2007ch}, or, including field fluctuations \cite{Brandenberger:2016egn, Ferreira:2016gfg}.}. The bulk spacetime is then a FRW-AdS and such a background can be conformally mapped to the familiar Schwarzschild-AdS black hole as analyzed in \cite{Apostolopoulos:2008ru}.

%\section{Model building}
{\it Model building} --
We begin with a holographic superconductor model with the action \cite{Franco:2009yz, Gubser:2008px, Hartnoll:2008kx, Hartnoll:2008vx} in form of
\begin{align}
 S = &\int d^5x\sqrt{-g} \big[ -\frac{1}{4}F_{MN}F^{MN} -\frac{1}{2}\left(\pd\Ps\right)^2 -\frac{1}{2}m^2\Ps^2 \no
 & -\Ps^n\(\pd_Mp-A_M\)\(\pd^Mp-A^M\) \big],
\end{align}
where $F_{MN}=\pd_M A_N-\pd_N A_M$ is the field strength of the bulk $U(1)$ gauge field. We use a complex scalar $\Ps$ as dual to the inflaton and the gauge field $A_M$ as dual to (fermionic) matter field by assuming that the inflaton decays to fermionic fields only. The fermionic fields could affect the cosmic expansion through backreaction.
In order to focus on the holographic description of the strongly coupled preheating, we do not consider backreaction in the present study.
The model is realized in an AdS-FRW background, which can be applied to describe our universe. In the probe limit, this background can be found from a coordinate transformation on an AdS-Schwarzschild black hole \cite{Apostolopoulos:2008ru}.
This offers a simplification, i.e.: as a first step we can study the dynamics of $\Ph$ and $A_M$ in an AdS-Schwarzschild black hole. With a conformal transformation, the results can be translated to the dynamics of inflaton and matter field in an expanding universe.

We consider a $(d+1)$-D AdS-Schwarzschild black hole background as follows,
\begin{align}\label{AdS_Schw}
 ds^2 = -f(r)dt^2 +\frac{dr^2}{f(r)}+\frac{r^2}{L^2}dx_i^2 ~,
\end{align}
where $f(r)=\frac{r^2}{L^2}-\frac{M^{d-2}}{r^{d-2}}$ and $i=1,2,\ldots, d-1$.
The horizon is given by $r_H \equiv M^{(d-2)/d}$. In the probe limit, the temperature is fixed by the horizon radius as:
%\begin{align}
 $T = \frac{r_H d}{4\pi L^2}$.
%\end{align}
From now on we set the AdS radius $L\equiv 1$ for simplicity.
To find the metastable background, we turn on both $\Ps$ and $A_t\equiv\Ph$. The dynamics of $\Ps$ and $\Ph$ can be obtained by solving the bulk equation of motion (EoM).
Near the AdS boundary $r\to \infty$, we get
\begin{align}\label{PsPh}
 &\Ps = \Ps_+ r^{-\Delta_+} +\Ps_- r^{-\Delta_-} +\cdots, \no
 &\Ph = \m -\frac{\r }{r^{d-2}} +\cdots ,
\end{align}
with
%\begin{align}
 $\Delta_\pm = \frac{d\pm\sqrt{d^2+4m^2}}{2}$.
%\end{align}

To apply to cosmic preheating, we set $d=4$ and $m^2=-3$ such that $\Delta_+=3$ and $\Delta_-=1$.
%In this case Eq.~\eqref{PsPh} contains additional logarithmic terms but they are irrelevant with background dynamics.
%We choose $m^2=-3$ (satisfying the Breitenlohner-Freedman bound in AdS$_5$) such that $\Delta_+=3$ and $\Delta_-=1$.
Both $\Ps_+$ and $\Ps_-$ are normalizable modes.
We take the quantization by adopting the Dirichlet boundary condition with $\Ps_+$ as the vacuum expectation value ({\it vev}) and $\Ps_-$ the source.
Thus the conformal dimension of the inflaton is $\Delta_+$.
Moreover, the boundary value of $\Ph$ gives the chemical potential $\m$ for the fermionic matter.
In analogy to holographic superconductors, there is a phase transition at a critical chemical potential $\m_c$, i.e., above $\m_c$ the phase is characterized by an inflaton condensate, while, below $\m_c$ the phase is normal matter with a vanishing {\it vev} of the inflaton.
%In analogy to holographic superconductors, there is a phase transition at a critical temperature $T_c$, i.e., below $T_c$ the phase is a hairy black hole, characterized by an inflaton condensate, while, above $T_c$ the phase is a hairless black hole, which is matter dominated with a vanishing {\it vev} of the inflaton.
The order of the phase transition depends on the choice of $n$, namely, it is second order for $n=2$ and first order for $n \ge 3$ \cite{Franco:2009yz}.
We choose the case of $n=3$, in which a metastable hairy black hole is present. %and the reason for this choice will be explained later.
The bulk EoM yields two possible solutions. The first is analytic:
%We proceed to solve Eq.~\eqref{eom_bg}, finding two possible solutions. One solution is analytic:
%\begin{align}\label{uncondensed}
 $\Ph = \m\(1-r_H^2/r^2\)$ and $\Ps=0$,
%\end{align}
where $\Ps$ is trivial and corresponds to a normal, uncondensed phase;
the second is solved numerically, which gives a vanishing source for the inflaton $\Ps_-=0$ but a non-vanishing {\it vev} $\Ps_+\ne 0$.

\begin{figure}
\begin{center}
\includegraphics[width=0.4\textwidth]{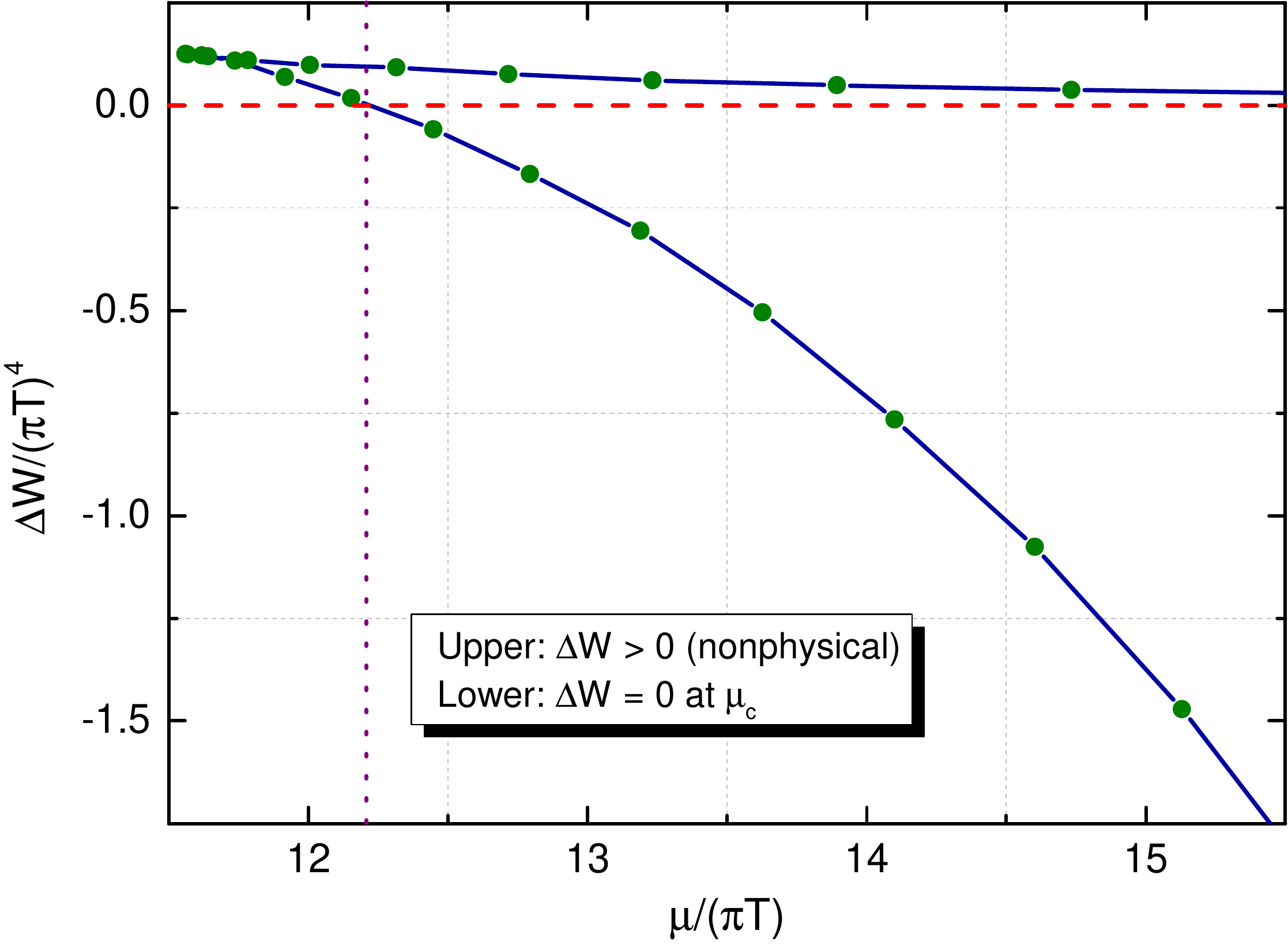}
\end{center}
\caption{The dimensionless free energy density difference $\D W$ as a function of dimensionless chemical potential $\m$ (normalized by $\pi T$). Model parameters are chosen as: $d=4$, $m^2=-3$ and $n=3$ in detailed numerics.
The background temperature is fixed at $T=0.2$, which is related to the energy scale of the system.
}\label{phase_diag}
\end{figure}

To obtain the phase diagram, we calculate the free energy associated with two solutions.
The free energy per unit volume for a general solution is given by,
%\begin{align}\label{free}
 $W = -\m\r-r_H^4\Ps_ +\Ps_- -\frac{3}{2} \int_{r_H}^\infty \frac{\Psi^3\Phi^2rdr}{1-r_H^4/r^4}$ \cite{Franco:2009yz}.
%\end{align}
For the first uncondensed solution, the free energy density takes $W_0 = -\m^2r_H^2$.
Thus, the difference of free energy density between the condensed and uncondensed phases is ,
\begin{align}\label{DeltaW}
 \D W = W -W_0 = \m^2 r_H^2 -\m\r - \frac{3}{2} \int_{r_H}^\infty \frac{\Psi^3\Phi^2rdr}{1-r_H^4/r^4} ,
\end{align}
where we have set the source of the bulk scalar $\Ps_-=0$.
We numerically plot $\D W$ as a function of $\m$ in Fig.~\ref{phase_diag}.
The uncondensed phase corresponds to $\D W=0$.
There exist two branches of the condensed phase as shown in Fig.~\ref{phase_diag}, of which the upper always has $\D W>0$ and thus is unphysical. The lower branch is stable ($\D W<0$) above a critical $\m_c$, and has a metastable ($\D W>0$) region when $\m < \m_c$, indicating a first order phase transition.
The metastable phase has a large {\it vev} of the inflaton.
%Interestingly we note that, at a sufficiently low temperature there may be more than one metastable states \cite{Franco:2009yz}, which greatly enriches the phase diagram but is out of our current scope.
%We will not consider this possibility and use the case of single metastable phase to model the initial state of preheating.
%
%focus on the beginning stage of the evolution, which is governed by the QNMs. The full analysis of the late time evolution will be addressed in our forthcoming work.
%
%The phase transition is seen as a first order one, with the condensed phase (inflaton dominated) extends below the critical chemical potential $\m_c$ \footnote{For a sufficiently low background temperature, there may exist more than one metastable states. We will not consider this possibility in this Letter.}.

We choose a metastable phase as the initial state.
As the inflaton dominated phase is thermodynamically unstable, the cosmic system will evolve towards the uncondensed phase, which is dominated by the matter field.
The detailed evolution of this process can be described by the QNMs of the background.
To this end, we consider the following perturbations:
$A_t\to \Phi+a_t$, $A_x\to a_x$, $\Psi\to \Psi+\sigma$ and $p\to\eta/\Psi$, with $a_t$, $a_x$ dual to the density and current of the produced matter and $\sigma$ dual to the inflaton condensate drained.
The details will be presented in an accompanied study \cite{Cai:2016b}. Here we quote the results, which are generic from the perspective of black hole QNMs and are crucial for cosmological implications.
The QNMs yield a dispersion relation $\o = \o(k)$ for a plane wave $e^{-i\o t+ikx}$ fluctuation, among which there is one mode containing positive imaginary part. This corresponds to an exponential growing mode associated with the instability of the metastable black hole;
the real part of the QNM vanishes in the homogeneous limit $k=0$, indicating a purely growing mode. A numerically small real part does show up for inhomogeneous fluctuations $k\ne0$.
Also, the dependence of QNMs on the inflaton's {\it vev} reveals that a large {\it vev} tends to trigger a fast conversion to matter field with the imaginary part being approximately linear in the {\it vev} \cite{Cai:2016b}.

%\section{Cosmological implications}
{\it Cosmological implications} --
Now we conformally map the above results to the AdS-FRW background to study cosmological implications.
It can be derived from the AdS-Schwarzschild black hole via coordinate transformations \cite{Apostolopoulos:2008ru}.
To be explicit, the 5D AdS-FRW in terms of the Fefferman-Graham coordinates takes,
\begin{eqnarray}\label{AdS_FRW}
 ds^2 = \frac{1}{\xi^2} \big[ d\xi^2 -\cN(\t,\xi)^2d\t^2 +\cA(\t,\xi)^2 d\vec{x}^2 \big] ~,
\end{eqnarray}
where $d\vec{x}^2 \equiv (dx_1^2+dx_2^2+dx_3^2)$.
The boundary of AdS-FRW is located at $\xi \to 0$ with $\cN\to1$ and $\cA\to a(\t)$, which corresponds to a 4D FRW universe. $a(\t)$ is the scale factor, governing the cosmic expansion.
The explicit forms of $\cN$ and $\cA$ can be expressed as \cite{Apostolopoulos:2008ru}:
%\begin{align}\label{cAN}
 $\cA^2 = a^2 -\frac{{\dot a}^2}{2}\xi^2 +\frac{({\dot a}^4 +4r_H^4)}{16a^2}\xi^4$, $\cN = \frac{{\dot \cA}}{{\dot a}}$,
%\end{align}
with the dot being derivative with respect to $\t$.
These two backgrounds are related via the following coordinate transformations,
\begin{align}\label{coord_trans}
 r = {\cA}/{\xi} ~,~ {\dot t} = -{({\dot \cA}r')}/{(f{\dot a})} ~,~ t' = -{{\dot a}}/{(\xi f)} ~,
\end{align}
with the prime being the derivative with respect to $\xi$.
The form of $t(\t, \xi)$ can be obtained by integrating \eqref{coord_trans}.
The coordinate transformations reduce to the conformal transformation on the boundary, which relates the energy stress tensor in FRW to that in Minkowski \cite{Apostolopoulos:2008ru}.
%We can derive analogous relations for the inflaton and matter field. The goal is to transform the metastable state and associated QNMs from the AdS-Schwarzschild to the AdS-FRW.

We start with the AdS-Schwarzschild background, with only $A_t=\Ph$ and $\Ps$ non-vanishing.
To apply holography, we consider the AdS-FRW background in the axial gauge with $A_\xi=0$. Then we perform the coordinate transformations and a gauge tranformation with
%\begin{align}
 $A_\xi = \(A_t+\pd_t\L\)t' +\pd_r\L r' .$
%\end{align}
The gauge condition $A_\xi=0$ fixes $\L(\t,\,\xi)$ as
%\begin{align}
$\L'=-A_tt'$.%$\sim O(\xi).$
%\end{align}
With a non-vanishing $\L$, $p$ gains a non-vanishing {\it vev} $p(\t,\,\xi)=\L$ in the AdS-FRW.
We further consider the QNMs parametrized by $\h$, $\s$, $a_t$ and $a_x$ as analyzed in \cite{Cai:2016b}.
Again, we impose the axial gauge condition $A_\xi+a_\xi=0$. It requires an additional gauge transformation with the parameter $\d\L$ satisfying
%\begin{align}
 $a_{\xi} = \(a_t+\pd_t\d\L\)t' +\pd_r\d\L r' =0 ,$
%\end{align}
which fixes:
%The gauge condition $A_\xi+a_\xi=0$ fixes $\d\L$ as
%\begin{align}\label{dL}
 $\d\L'=-a_tt'$.
%\end{align}
%
%We are one step away from obtaining the QNMs in FRW-AdS background.

Afterwards, we are ready to derive the QNMs in the AdS-FRW by applying the previous coordinate and gauge transformations. We focus on the $a_t$ component that is related to the matter field creation. Parametrizing the form of $a_t$ as
\begin{align}\label{qnm_at}
 a_t = e^{-i\o t+ikx} F_t(r), \text{with}\;\; F_t(r) = \frac{\#}{r^2} +\cdots , \nonumber
\end{align}
where $\#$ is a numerical prefactor that encodes the dependence on the inflaton's {\it vev}. Using asymptotic forms of coordinate transformations,
%\begin{align}
% ${\dot t} = \frac{1}{a}+ ( \frac{{\dot a}^2}{a^3}-\frac{{\ddot a}}{2a^2} ) \xi^2+O(\xi^3)$,
% $t' = -\frac{{\dot a}}{a^2}\xi +O(\xi^3)$,
% $r =\frac{a}{\xi} +O(\xi)$,
%\end{align}
one gets
\begin{align}
 a_t %= e^{[ -i\(\int\frac{\o d\t}{a(\t)} -\frac{{\dot a}}{2a^2}\xi^2\) +\cdots+ikx ]} \big(\frac{\#\xi^2}{a^2} +\cdots \big) \no
 \simeq e^{[ -i\o\int\frac{ d\t}{a(\t)} +ikx ]} \Big( \frac{\#\xi^2}{a^2} +O(\xi^3) \Big) ,
\end{align}
and hence, $a_tt'\sim O(\xi^3)$, $\d\L\sim O(\xi^4)$. The QNM $a_\t$ in the AdS-FRW background takes
%\begin{align}
 $a_\t = \(a_t +\pd_t\d\L\){\dot t} +\pd_r\d\L{\dot r} = a_t{\dot t} +{\dot {\d\L}} ~$.
%\end{align}
From the gauge condition, $\d\L\sim O(\xi^4)$, we find that it is negligible in contributing to matter field. Eventually, one gets
\begin{align}\label{qnm_frw}
a_\t = e^{ [ -i\o\int\frac{d\t}{a(\t)} +ikx ] } \big( \frac{\#\xi^2}{a^3} +\cdots \big)
\end{align}
in the AdS-FRW.
Using the dictionary derived in \cite{Cai:2016b}, one obtains the growth of matter density as
\begin{align}\label{jt}
 j^\t = e^{\(-i\int\frac{\o d\t}{a(\t)}+ikx\)}\frac{\#}{a^3} ~.
\end{align}

The structure of \eqref{jt} is very instructive. It consists of an exponent and an overall factor.
The scale factor $a(\t)$ enters both in the exponent and the overall factor. In the exponent, it appears through the ``rescaled time'' $\int {d\t}/{a(\t)}$. As we have discussed before, QNMs with the positive imaginary part correspond to the existence of an instability, which grows according to the rescaled time.
An expanding universe would typically suppress the instability and thus slow down the amplification of matter field.
On the other hand, the overall factor ${1}/{a^3}$ depicts the damping effect due to the cosmic expansion.
Additionally, the prefactor $\#$ encodes the dependence on the inflaton's {\it vev}, as is inherited from the Minkowski result.

Eventually, we compare the results of holographic preheating at strong coupling with those obtained at weak coupling. Here we quote the results from a comprehensive review \cite{Allahverdi:2010xz} (adapted in our notations)
\begin{align}\label{weak}
 n(\t) \sim \frac{(m\Ph_0)^{3/2} e^{2m\m \t}}{a^3 \sqrt{1+ \frac{m\m \t}{2\pi} }} ~,
\end{align}
where $\m$ and $\Ph_0$ are the mass and {\it vev} of inflaton. In the weak coupling case, the inflaton field oscillates in the potential and produces matter particles every time it passes the bottom, which corresponds to the moment when the adiabaticity condition breaks down. We note several interesting properties between \eqref{jt} and \eqref{weak} as follows: \\
(i) Unlike in the weak coupling case with particles being produced discontinuously, the strong coupling scenario is featured by continuous particle production. Oscillation is only seen in the amplification of inhomogeneous matter density in the strong coupling case. \\
(ii) Both two share the same overall factor ${1}/{a^3}$, which is due to the damping effect from the cosmic expansion.
%It is tempting to conjecture that the factor is generic for any homogeneous density.
We stress that the damping effect at strong coupling is nonadiabatic as $a$ also appears in the rescaled time. \\
(iii) The qualitative dependence of the growth rate on the inflaton's {\it vev} is the same: a large {\it vev} leads to a rapid growth. But the details are different.
The strong coupling result \eqref{jt} gives a stronger dependence on the inflaton's {\it vev}, i.e. the exponent of matter growth/inflaton depletion grows with the inflaton's {\it vev}; while, in the weak coupling case \eqref{weak}, the inflaton's {\it vev} appears merely in a power law form.

%\section{Conclusion and Outlook}
{\it Conclusions.--}
In this Letter we proposed a brand new description of cosmic preheating at strong coupling by virtue of a holographic dictionary.
%have studied the cosmological preheating process by virtue of a holographic model, dual to strongly coupled inflaton and matter field.
Our results are different from those obtained in the weak coupling case.
The most distinguished difference is that in the strongly coupled case, the matter field can be amplified continuously in contrast to the weakly coupled case where the matter field is only produced when the parametric resonance is triggered.
%the inflaton passes over the bottom of its potential.
The scenario of {\it holographic preheating} is expected to be generic from the viewpoint of the QNMs in a metastable black hole background.
%Thus, we argue that this qualitative feature will not be much changed in different holographic models.

The proposed scenario may initiate fruitful studies from many perspectives.
Note that, we modelled a specific initial state with a metastable black hole in the probe limit, which can be generalized.
Also, we assumed that the inflaton only decays to fermionic matter, which can be extended to the bosonic case.
The analysis of cosmic preheating should include the backreaction, which under holographic description corresponds to the evolution from a metastable hairy black hole to a stable hairless black hole.
The end of this evolution is dual to the matter field dominated phase of the universe.
While the process of equilibration is hard to be realized in the weak coupling case \cite{Micha:2004bv, Micha:2002ey}, it becomes natural from the holographic point of view of the black hole dynamics.
These issues are left for forthcoming study.

We also interestingly note that, the {\it holographic preheating} that describes preheating at strong coupling may be applied to inflationary models from  fundamental theories, such as, string theory. With the proposed scenario, we could better understand the universe with strong coupling and further discriminate various models of the very early universe by combining with cosmological observations.

%\section*{Acknowledgments}
{\it Acknowledgments.--}
We are grateful to R. Brandenberger, S. Carroll, Y. Hu, H. Lu, A. Marciano, R.-X. Miao, R. Niu, W.-Y. Wen and Y. Yang for valuable communications.
YFC is supported in part by the Chinese National Youth Thousand Talents Program (No. KJ2030220006), by the USTC start-up (No. KY2030000049) and by NSFC (Nos. 11421303, 11653002).
SL is supported in part by the Chinese National Youth Thousand Talents Program and by the SYSU Junior Faculty Fund.
JL is supported in part by the physics department at Caltech and by the Fund for Fostering Talents in Basic Science of NSFC (No. J1310021).
JRS is supported in part by NSFC (No. 11205058), by the Fundamental Research Funds for Central Universities and by the Open Project Program of State Key Laboratory of Theoretical Physics at CAS (No. Y5KF161CJ1).
Part of numerics are operated on the computer cluster LINDA in the particle cosmology group at USTC.

\end{document}